# Applications of nanoparticles of γ $Fe_2O_3$ for hyperthermia in *E.coli* by Nd:YAG laser

Srikanya Kundu[1], Harshada Nagar[1], S D Kulkarni[2], Renu Pasricha[2], A K Das[3], G R Kulkarni[1] and S V Bhoraskar[1, *]

[1] *Department of Physics, University of Pune, Pune – 411007, India,* [2] *Laser and Plasma Technology Division, BARC, Mumbai – 400085, India,* [3] *National Chemical Laboratory, Pashan Road, Pune- 411 008, India,*

[*] *Author for correspondence (Tel.: +91-020-25692678 (ext: 427); FAX: +91-020-25691684; Email: svb@physics.unipune.ernet.in)*

**Abstract**

The paper explores the use of nanoparticles of γ $Fe_2O_3$ for hyperthermia treatment of living organisms by absorption of 1064 nm radiations from Nd:YAG laser. *Escherichia coli* cells have been used as the model system for demonstrating the effect wherein lysine is used as an interface between the cell walls and the nanoparticles. Scanning Electron Microscopic observations have, exclusively, proved that attachment of nanoparticles of iron oxide along with lysine alone is responsible for absorption of above radiations. The quantitative estimation has been provided by growth rate measurements and protein assessment of the cells. The nanoparticles of γ $Fe_2O_3$ were synthesized by DC arc plasma assisted gas phase condensation.

**Introduction**

Applications of nanosized particles for *in vitro* diagnostics have been practiced for several years. However its use is outstandingly being recognized only recently in biomedical therapeutic fields. Nanoparticles, on account of their large surface to volume ratio offer additional advantages of properties like tissue accessibility due to the suitable interfacing. This helps the nanoparticles to play a vital role in several biomedical applications such as drug delivery (Berry et al., 2003), hyperthermia (Jordan et al., 1999; Gonzales et al., 2005), cellular therapy (Vries et al., 2005; Rogers et al., 2006) , targeting for cell biology research (Gupta & Gupta, 2005), and tissue repair. A potential benefit of using magnetic nanoparticles is in the field of localized magnetic field gradients. The nanoparticles can be injected intravenously and transported through blood stream to the desired area of treatment (Berry et al., 2003).

In most cases super paramagnetic particles (with sizes ranging from 10 – 20 nm) are of greater interest for *in vivo* applications. As they do not retain any magnetism after removal of magnetic field and they are physiologically well tolerated. There are reports where magnetic nanoparticles are deposited on tumor tissues and are heated in an alternating magnetic field in order to destroy the tumor (Forbes et al., 2003; Muller et al., 2005).

This paper explores the feasibility of using nanocrystalline γ $Fe_2O_3$ for destroying the living bioorganisms by hyperthermia effect, which involves heating of certain organs or tissues to facilitate their destruction. Iron oxide exhibits fairly strong absorption for infrared radiations at 1064 nm from Nd:YAG laser. The thermal radiations are subsequently transferred to cell walls and thus produce the damage. Negligible absorption of Nd:YAG (λ = 1064 nm) laser by the biological tissues makes it suitable for therapeutic applications with minimum side effects. However no such attempt has been so far reported to combine the laser irradiation and the destruction of bioorganisms by making use of nanoparticles.

On account of the simplicity of experimentation *E.coli* (wild strain) has been used as a living organism. The nanoparticles of γ $Fe_2O_3$ ranging in size between 5 – 20 nm have been synthesized by the process of gas phase condensation in an arc plasma device. The lysis of cell walls by the hyperthermia treatment induced by Nd:YAG laser was investigated by scanning electron microscopic observations as well as by monitoring the growth rate of bacteria. After lysis of the cell walls the protein released from the microorganism was estimated by Bradford protein assay.

**Materials and methods**

Nanocrystalline gamma ferric oxide was synthesized by dc arc- plasma assisted gas phase condensation. The system consisted of an arc initiated thermal- plasma reactor in which iron block was used as an anode and graphite was the cathode. The schematic of the experimental set up is shown in Figure1. The morphology of the particles was investigated using Transmission Electron Microscope with JEOL model 1200 EX.

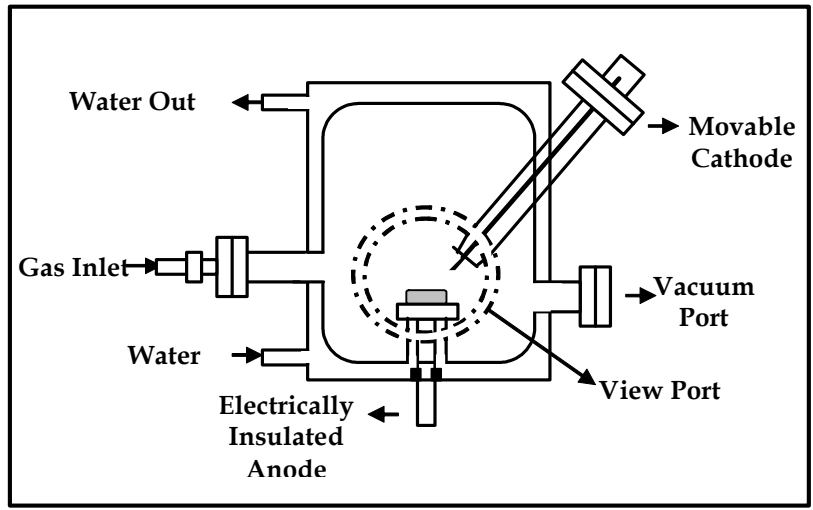

*Figure 1*: Schematic of the experimental set up of nanoparticles synthesis

Amino acid lysine was coated on the particles so as to create an electrostatic interaction between the negatively charged *E.coli* cell membrane and the iron oxide particles. Lysine (Hydropathy index -3.9) (Nelson & Cox, 2005) and iron oxide were mixed in 2:1 proportion in 100 ml of distilled water (Stanislav et al., 2002). The medium was sonicated for 45 minutes at room temperature.

*E.coli* cells were cultured in a flask with a Nutrient broth medium (peptone-3.45 gm, beaf extract-3.45 gm, NaCl-5.1 gm, distilled water-1000 ml, pH- 7.5), This culture was incubated for 16-18 hours at $37^0$C. Optical density (OD) was measured at λ=600 nm by using SHIMADZU UV-1601 Spectrophotometer. The cell concentration was estimated to be $10^7$ cells/ml. The cells were separated from the solution by centrifuging 15 ml of culture at 4500 rpm for 10 min at room temperature. The live cells in the culture were subsequently washed with 1 (M) NaOH solution for cleaning the extracellular surface protein substances (EPS) and also to expose the largest possible amount of negatively charged peptidoglycan surface of *E.coli* cell membrane (Berry et al., 2004). After washing the cells they were again subjected to centrifuge action at a speed of 4500 rpm at room temperature and the cell palettes were collected. Subsequently the product was dissolved in 5ml of sterile distilled water. The washed bacterial suspension, so obtained, was exposed to 1 ml of lysine coated γ-$Fe_2O_3$ suspension having a pH value of 7 for 16 hours with continuous gentle stirring.

Figure 2 shows a flow chart of the complete process. The deposition of subsequent layers of positively charged γ-$Fe_2O_3$ is self inhibitory due to particle – particle charge based repulsion. Bacterial cell viability was maintained during complete experimental procedure. The strong specific affinity of the bacterial cell membrane (electronegative) to lysine was critical for achieving area coverage over the percolation threshold of 45 % for two dimensional structures (Berry et al., 2004).

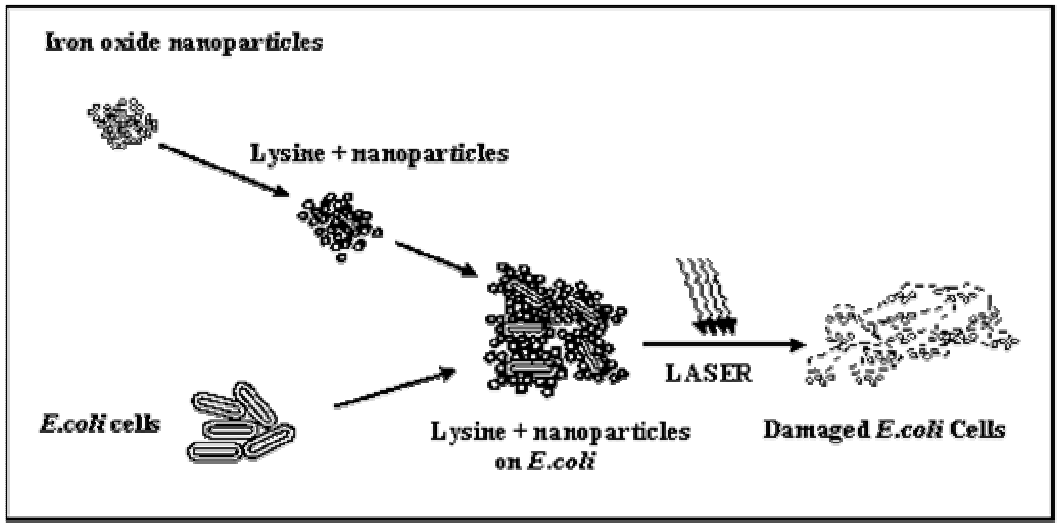

*Figure 2.* Schematic representation of the experimental procedure

Nd:YAG laser with Continuous Wave (CW) mode and variable power having λ = 1064 nm was used for irradiation. *E.coli* cells were irradiated for 1 minute at two different laser powers. Prior to the irradiation the cultures (in Nutrient broth) were centrifuged and the cells were removed from water and subsequently irradiated by laser. The laser irradiated *E.coli* cells were then suspended in sterile distilled water.

The irradiated as well as unirradiated cell suspensions were divided into two sets. One set of each was used for investigations under Scanning Electron Microscopy (SEM) and Energy Dispersive X-ray Analysis (EDXA) (JEOL JSM-6360A, Analytical Scanning Electron Microscope) where as the other set was subjected to the growth rate measurement. First set was again centrifuged at 4500 rpm for 10 minutes to separate the supernatant and the palette of the cells. Standard Bradford protein assay were carried out to estimate the amount of protein released from lysised cells due to hyperthermia which shows a linear relationship with the extent of celldestruction (Bradford, 1976).

The later set of experiments was carried out by inoculating 1 ml of *E.coli* in 100 ml of nutrient broth and incubating at $37^0$C while stirring 2 ml of aliquots were collected from the broth after every 40 minutes intervals (20 min *E.coli* growth cycles, two generations). The optical absorbance was measured for each of these samples at 600 nm using a SHIMADZU UV-1601 Spectrophotometer. This provided the measurement for increase in turbidity of the medium, which is related to the growth rate of bacteria (Black, 1996). These measurements were carried out for irradiated as well as unirradiated *E.coli*. The collection was continued for 7 hours. Growth curves of coated and uncoated *E.coli* samples with and without exposure to Nd:YAG laser radiation were plotted on the basis of the turbidometric data.

**Results and discussions**

The morphology of the as synthesized powder was analyzed with the help of Transmission Electron Microscopy (TEM) micrographs. The particles were spherical in shape and varied in size from 15 to 80 nm. This wide size distribution is common for the gas phase synthesis process. The crystalline phase was identified by X-ray diffraction analysis and was found to be $\gamma Fe_2O_3$ as reported in our earlier communication (Banerjee et al., 2007). The magnetic behavior of $\gamma Fe_2O_3$ was investigated by magnetization curve by vibrating sample magnetometer. Figure 3 illustrates all the important characteristic properties of the precursor $\gamma Fe_2O_3$ used in the present studies.

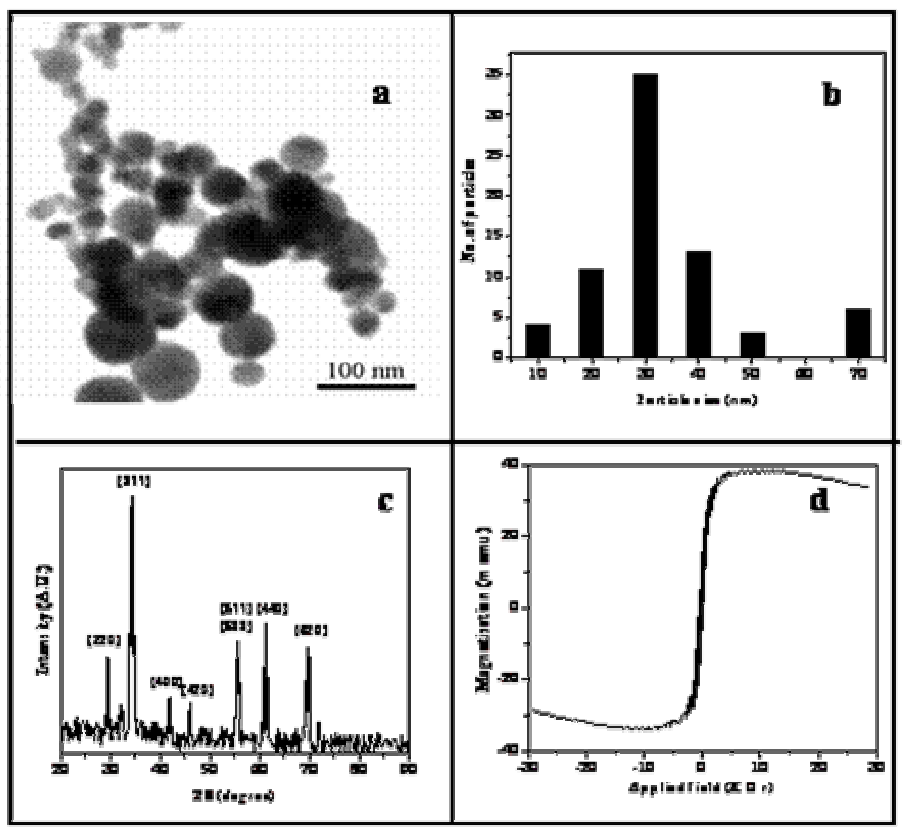

*Figure 3*. Characterization of as synthesized powder of iron oxide

a) Transmission electron micrograph b) The corresponding size histogram c) X ray diffraction spectrum of $F_2O_3$ and d) Magnetisation curve recorded by vibrating sample magnetometer

The particle size histogram obtained from large number of sample analysis provides a feature indicated by the histogram shown in Figure 3 (b). The wide particle size distribution is preferred while coating the microorganism since then the particles can easily get accommodated on their irregular shapes and surfaces.

In order to ensure the hyperthermia reaction, the rise in temperature of the coated cells under irradiation was monitored in a separate experiment by making use of a temperature sensor. The measured temperatures were plotted in the bar diagram shown in Figure 4.

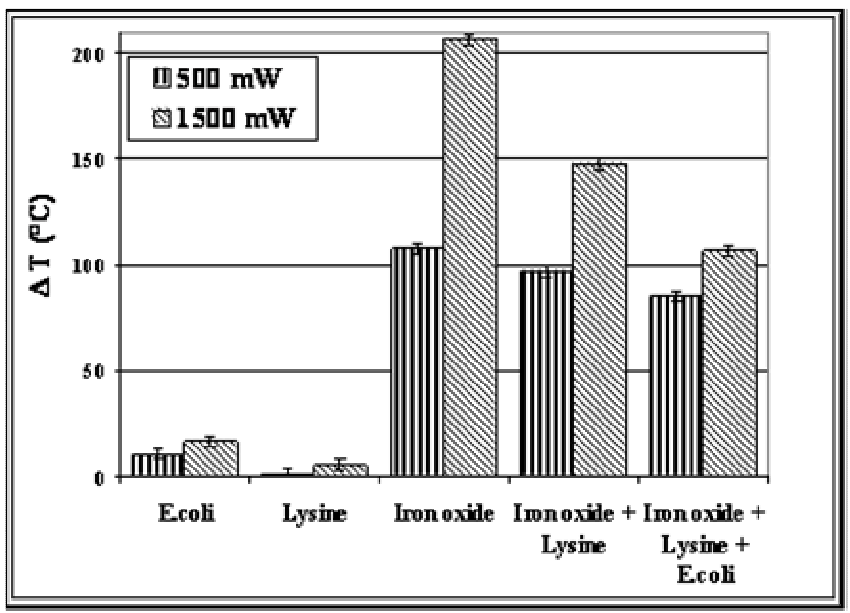

*Figure 4*: Bar diagram showing the rise in temperature ΔT (°C) for different experimental systems when exposed to laser beam and measured with a temperature sensor

*E.coli* and lysine alone did not show appreciable rise in temperature, where in the *E.coli* cells when coated with lysine and iron oxide provided sufficient increase in the temperature when exposed to laser radiations.

Nanoparticles of $Fe_2O_3$ were deposited on the *E.coli* cells as discussed in the experimental section and the SEM images were recorded before and after exposure to the laser irradiation at two different powers. Figure 5 and Figure 6 exhibit the morphological changes seen in *E.coli* during the experiments.

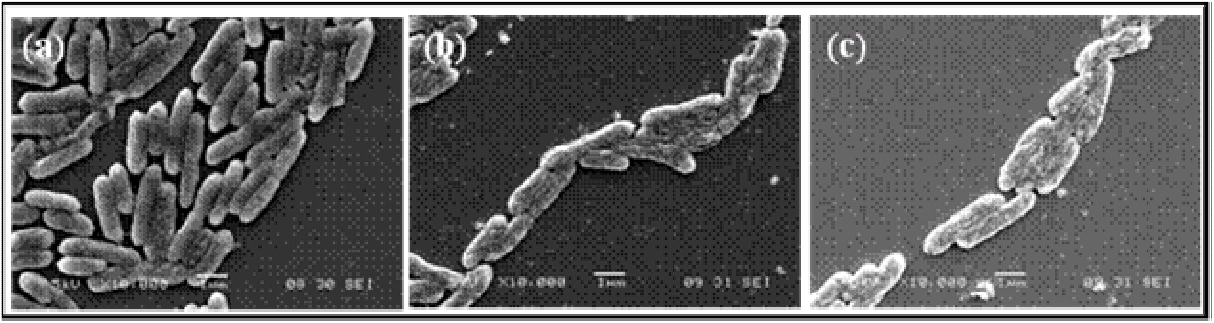

*Figure 5*. SEM pictures of the *E.coli* (a) Virgin (b) Irradiated with Nd:YAG at 500 mW

(c) irradiated with 1500 mW

Figure 5 (a) shows a micrograph of *E.coli* alone, showing the separated cells having dimensions of roughly 0.5 µm diameter and 2 µm length. Figure 5 (b) shows the image of the same when irradiated with 500 mW laser power and Figure 5 (c) when irradiated with 1500 mW laser power. The *E.coli* cells, in these two micrographs are seen to be almost unaffected except that they get assembled and the surface roughness of the cells.

These micrographs can be compared with Figure 6 for the coated *E.coli*. Figure 6 (a) shows the micrographs of the *E.coli*. coated with iron oxide. Increased agglomeration on the cell surface can be easily visualized.

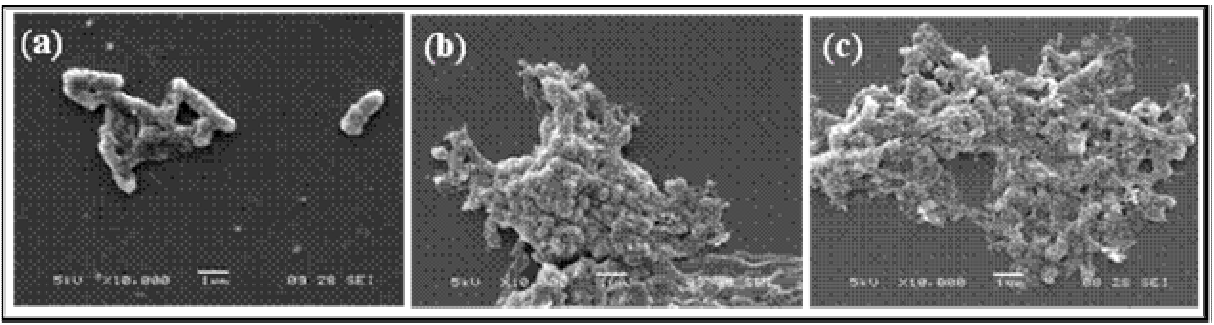

*Figure 6*. SEM images of the *E.coli* (a) coated by nano $Fe_2O_3$ (b) coated and irradiated by

1500 mW (c) coated and irradiated with 1500 mW

Figure 6 (b) projects the image of these iron oxide coated *E.coli*. cells after irradiation with 500 mW. Similarly Figure 6 (c) shows the effect of irradiation of the coated cells with 1500 mW. Cell lysis is observed, unambiguously, in the last two images. The extent of lysis has increased with increasing laser power.

Consecutively the quantitative estimation of cell lysis due to laser irradiation for both the laser powers was monitored by release of cellular proteins through Bradford protein estimation. A standard protein curve was first plotted for Bovine Serum Albumin (BSA). The absorbance values for the irradiated cell suspensions were then experimentally measured at 595 nm and the corresponding protein concentrations were estimated by extrapolating straight line as indicated in Figure 7.

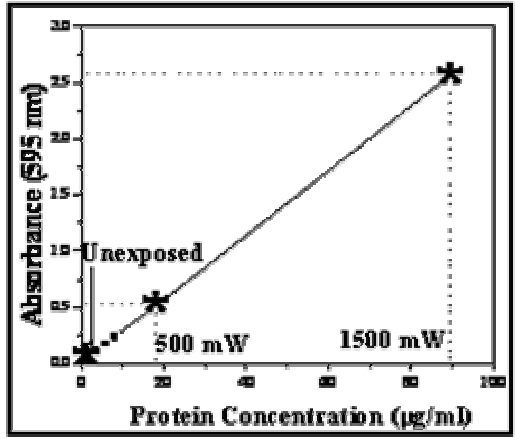

*Figure 7*. Protein estimation curve of unirradiated and irradiated *E.coli* cells for 500 mW and 1500 mWlaser power are indicated

The protein contents were found to be 1.6 µg/ml for the unirradiated and 18.41 µg/ml and 89.81 µg/ml for the irradiated cells, respectively for the two laser powers. The increase in the protein content with varying laser power clearly indicates the extent of cell damage caused by the thermal effects.

In order to investigate the duration of thermal treatment effects caused by laser irradiation, we tried to estimate the growth rate by optical absorption measurements (Black, 1996). The absorbance at 600 nm has been plotted as a function of incubation time of the *E.coli* cells for the controlled as well as irradiated cells. This shows a sigmoidal pattern.

The estimation of growth rate as obtained by absorbance measurement is shown in Figure 8. This shows the sigmoidal pattern which is a characteristic of bacterial growth in nutrient rich medium. Figure 8 (b) shows the growth curve for $\gamma$-$Fe_2O_3$ coated *E.coli* exposed to Nd:YAG laser for 500 mW & 1500 mW laser powers exhibiting a slight deviation from the sigmoidal pattern.

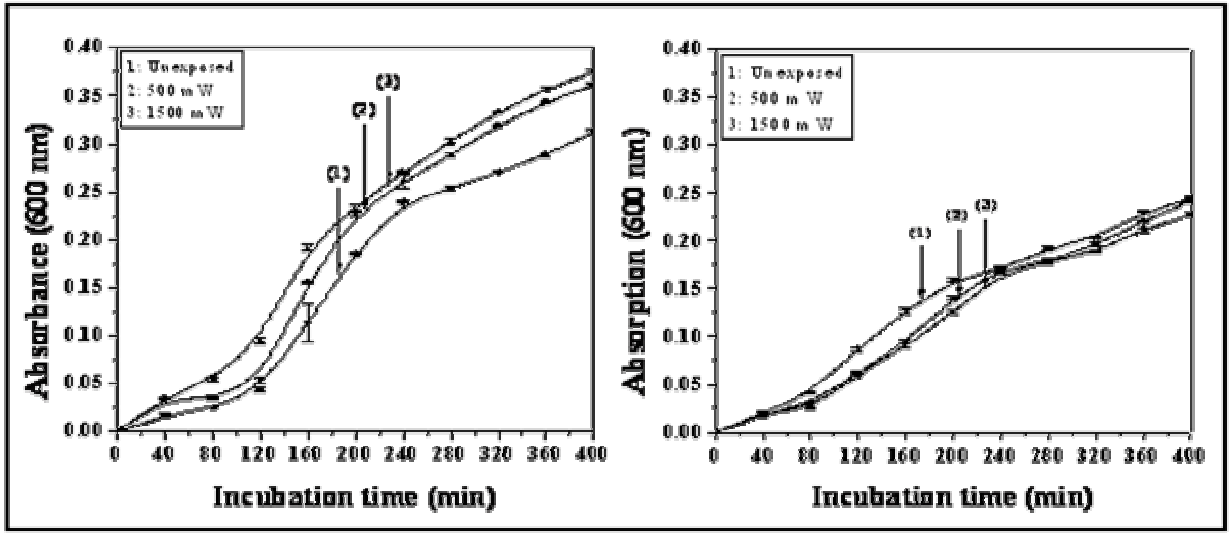

*Figure 8*. Growth curves for *E.coli* cells (a) Normal *E.coli* growth curves under different power of Nd:YAG laser exposure for one minute (b) $\gamma$-$Fe_2O_3$ coated *E.coli* growth curves under different power of Nd:YAG laser exposure for one minute

Growth rate curves provide additional quantitative evidence in support of the present experiments. It has been seen that by increasing the laser power, growth rate of the uncoated *E.coli* has increased by about 40%. On the other hand $\gamma$-$Fe_2O_3$ coated *E.coli* shows a significant decrease in the growth rate with respect to the uncoated *E.coli*. The overall growth has decreased by about 20%. The increase in the population of the uncoated *E.coli* due to the laser irradiation indicates the growth stimulation in presence of the radiations (Nussbaum et al., 2003). The decrease of growth rate in the coated cells after irradiation, in spite of the stimulating action of radiations on *E.coli*, therefore point out towards the cell lysis caused by the temperature rise due to laser absorbance of $\gamma$-$Fe_2O_3$ alone.

**Conclusions**

It is confirmed that nanoparticles of iron oxide destruct *E.coli* cells by the local heating effect due to laser radiations. It has been shown that nanoparticles alone are responsible for inhibiting the growth rate and imparting damage to the cells. Thus this process of hyperthermia using $Fe_2O_3$ nanoparticles and Nd:YAG laser can be extended to destroy the malignant cells with minimal side effects.

**Acknowledgement**

The authors wish to acknowledge DAE-BRNS and DRDO for the financial support.